\documentclass[preprint,prb]{revtex4}
\usepackage{amssymb}
\usepackage{graphicx}

\newcommand{\R}{{\mathbb R}}
\newcommand{\lP}{\ell_{{\rm P}}}

\begin{document}

\preprint{AEI--2003--114}

\title{Loop Quantum Cosmology and Boundary Proposals}

\author{Martin Bojowald\footnote{mabo@aei.mpg.de}
}
\affiliation{Max-Planck-Institute for Gravitational Physics, Albert Einstein Institute,\\
Am M\"uhlenberg 1, D--14476 Golm, Germany}

\author{Kevin Vandersloot\footnote{kfvander@gravity.psu.edu}}
\affiliation{Center for Gravitational Physics and Geometry,\\
104 Davey Lab, University Park, PA 16802, USA}

\begin{abstract}
For many years, the most active area of quantum cosmology
has been the issue of choosing boundary conditions for the wave
function of a universe. Recently, loop quantum cosmology, which is
obtained from loop quantum gravity, has shed new light on this
question. In this case, boundary conditions are not chosen by hand
with some particular physical intuition in mind, but they are part of
the dynamical law. It is then natural to ask if there are any
relations between these boundary conditions and the ones provided
before. After discussing the technical foundation of loop quantum
cosmology which leads to crucial differences to the Wheeler--DeWitt
quantization, we compare the dynamical initial conditions of loop
quantum cosmology with the tunneling and the no-boundary proposal and
explain why they are closer to the no-boundary condition. We end with
a discussion of recent developments and several open problems of loop
quantum cosmology.
\end{abstract}

\maketitle

\section{Introduction}

Loop quantum cosmology is the cosmological sector of loop quantum
gravity,\cite{Rov:Loops,ThomasRev} a canonical quantization of
general relativity using Ashtekar's variables. In this
formulation, gravity becomes a constrained gauge theory with the basic
classical fields being an SU(2)-connection $A_a^i$, a sum of the spin
connection and the extrinsic curvature, and its conjugate, the
densitized triad $E_i^a$. These fields are represented in the
quantum theory by holonomies of the connection and fluxes of the
triad, which allows a mathematically well-defines formulation of
the full theory.\cite{ALMMT} Nevertheless, the complexity of the
full formalism makes it advantageous to consider simplified models
in a minisuperspace quantization. Since there is a well-developed
mathematical structure behind the full theory, the reduction to
symmetric models can be done in such a way that essential features
of the full theory, e.g., the discreteness of geometric spectra,
are preserved. This is in contrast to the older Wheeler--DeWitt
quantization\cite{DeWitt,Misner,QCReview} where the full theory
would not even be known beyond a purely formal level.

Technically, states of the full theory are functions on the space
of connections which depend on $A_a^i$ only via holonomies
(cylindrical functions). Symmetric states are by definition
distributions on the space of connections which are supported only
on the sub-manifold of invariant connections.\cite{SymmRed} It
turns out that such a reduction even to the simplest cosmological
models yields a quantum representation which is inequivalent to
the one chosen in the Wheeler--DeWitt quantization (which is
possible also in systems with a finite number of degrees of
freedom since the Weyl algebra is not represented
continuously).\cite{Bohr} This translates into a different point
of view as to the representation of basic variables. While the
Wheeler--DeWitt quantization assumes that the extrinsic curvature
can be promoted to a well-defined operator, the background
independent techniques employed in the loop quantization allow
only a representation of holonomies as operators, not the
connection itself which in the isotropic setting corresponds to
the extrinsic curvature. Also triad operators behave differently:
they would have continuous spectra in a Wheeler--DeWitt
quantization while they have discrete spectra in a loop
quantization. On the other hand, these basic properties are the
same in full loop quantum gravity and in loop quantum cosmology
such that one can regard those cosmological models as reliable
tests.

These basic technical details have far-reaching physical
consequences. For instance, the Wheeler--DeWitt quantization cannot
eliminate the cosmological singularities that plague general
relativity, while they do not occur in loop quantum
cosmology.\cite{Sing} Also the issue of boundary conditions, which has
been the main issue in the Wheeler--DeWitt quantization, appears in a
different light.\cite{DynIn,Essay} All these effects are intimately
linked to the underlying discreteness of geometry.

In this contribution we describe both the basic formulation of the
theory, emphasizing differences to the Wheeler--DeWitt quantization,
and physical applications. Both parts can be read independently of
each other.

\section{Basic Formalism of Loop Quantum Gravity and Loop Quantum Cosmology}

Loop quantum gravity is based on general relativity in Ashtekar's
canonical variables\cite{AshVar,AshVarReell}
$A_a^i=\Gamma_a^i(E)-\gamma K^i_a$ and $E^a_i$ on a spacelike slice
$\Sigma$, where $\Gamma_a^i$ is the spin connection associated with
the densitized triad $E^a_i$, $K_a^i$ the extrinsic curvature, and
$\gamma$ the Barbero--Immirzi parameter which is a positive real
number. For a background independent quantization it is important to
represent these basic fields via holonomies
\[
 h_e(A)={\mathcal P}\exp\int_e A_a^i\tau_i \dot{e}^a dt
\]
where $e$ is a curve in $\Sigma$ with tangent vector field $\dot{e}^a$
and $\tau_i$ are Pauli matrices, and fluxes
\[
 F^{(f)}_S(E)=\int_S E^a_i f^i n_a d^2y
\]
where $S$ is a surface in $\Sigma$ with co-normal $n_a$ and $f^i$ a
test function. If all curves, surfaces and test functions are allowed,
these quantities have the full information contained in the fields
$A_a^i$ and $E^a_i$.

The variables are constrained by the usual SU(2)-Gauss constraint as
well as the diffeomorphism constraint, which requires invariance under
deformations of space, and the Hamiltonian constraint which yields the
dynamics of the theory. Since the diffeomorphisms act directly on the
basic fields, a representation of their classical Poisson-*-algebra
has to be diffeomorphism invariant if this classical symmetry is to be
preserved in the quantization.

\subsection{Representation of the Basic Classical Algebra}

There is in fact a natural diffeomorphism invariant representation
of the holonomy/flux algebra which has recently been shown to be
unique.\cite{FluxAlg,Meas,SuperSel,Irred,HolFluxRep} A heuristic
construction of this representation can be done as follows. Since
there is a rich mathematical structure on the space of
connections, it is convenient to look for the state space in the
connection representation. The simplest such state is the ``ground
state'' {\bf 1}$(A)=1$ which does not depend on the connection at
all. All higher states can then be ``created'' by acting with
holonomies as multiplication operators. Since one can act with
several holonomies, one at a time, the final states will depend on
the connection non-trivially, but only along the edges in the
holonomies chosen, and only via the holonomies (cylindrical
functions). Labels of a given state, the edges of the holonomies
used to construct it, can be arranged to a graph in $\Sigma$ which
may have self-intersections and also knotting and linking
information. Furthermore, since one can act with the same holonomy
several times, one also needs a label representing the number of
times a given edge occurs in the graph. Another way to include
this information is by using irreducible SU(2)-representations
since acting once with a holonomy creates dependence of the state
via the fundamental representation while acting several times then
creates dependence according to the SU(2)-coupling rules. An
independent set of states can be seen to be given by the spin
network basis\cite{RS:Spinnet} given by graphs whose edges are
labeled by irreducible SU(2)-representations, or half-integer
spin labels $j$.

While any such state depends on the connection only via finitely many
holonomies, the full state space is obtained as a projective limit
over the set of graphs.\cite{FuncInt,ALMMT} Thus, there is no
restriction on the dependence of states on connections and one obtains
a faithful representation of the infinite dimensional space of
connections. This state space carries a natural diffeomorphism
invariant inner product, the Ashtekar--Lewandowski inner product,
which defines the Hilbert space of loop quantum gravity.

Fluxes represent the conjugate of the connection and thus act as
(functional) derivative operators in the connection
representation. Since the states depend on the connection via
holonomies, the action of flux operators can be obtained from the
chain rule. A flux operator associated with a surface $S$ has a
non-trivial action on a state labeled by a graph $g$ only if $S$
and $g$ intersect each other transversally. If they do, the action
is given by a sum over transversal intersection points, each
contribution counting the multiplicity of the edge encoded in its
spin label $j$. Thus, flux operators count the intersection number
of the surface with the graph of a state, counting also
multiplicities, which is an integer. In this way, flux operators
obtain a discrete spectrum which translates to discrete spectra of
other geometrical operators constructed from the densitized triad,
such as area or volume.\cite{AreaVol,Area,Vol2}

\subsection{Cosmological Models}

In order to follow the reduction procedure, we need to know the
sub-manifold of invariant connections for an isotropic symmetry
group. Any isotropic connection can be written in the form\cite{cosmoI}
\[
 A_a^i=c\Lambda^i_I \omega_a^I
\]
where $\omega^I$ are left-invariant 1-forms on $\Sigma$ for the action
of the symmetry group, $\Lambda$ is an internal triad (an
$SO(3)$-matrix) which is purely gauge, and $c$ is the only
gauge-invariant component of the connection. Similarly, an isotropic
triad has the dual form
\[
 E^a_i=p\Lambda^I_i X^a_I
\]
with left-invariant vector fields $X_I$ dual to $\omega^I$.

Isotropic states are then distributions in the full state space which
are supported only on isotropic connections. Thus, gauge invariant
isotropic states can depend only on the component $c$, and an explicit
reduction, e.g.\ by restricting spin network states to isotropic
connections, shows that any isotropic state can be represented as
\[
 |\psi\rangle = \sum_{\mu} \psi_{\mu} |\mu\rangle
\]
where the sum is over a countable subset of $\R$. The isotropic states
\begin{equation} \label{states}
 \langle c|\mu\rangle = e^{i\mu c}
\end{equation}
form an orthonormal basis since their inner product is given by
\begin{equation} \label{inner}
 \langle\mu|\mu'\rangle = \delta_{\mu,\mu'}
\end{equation}
which implies that the Hilbert space is non-separable (all $\mu\in\R$
are allowed).

In this connection representation, holonomies $e^{i\mu c}$ act as
multiplication operators
\begin{equation} \label{c}
 e^{i\mu c}|\mu'\rangle = |\mu+\mu'\rangle
\end{equation}
and the flux $p$ as a derivative operator
\begin{equation} \label{p}
 \hat{p}|\mu\rangle = \frac{1}{6}\gamma\lP^2\mu|\mu\rangle\,.
\end{equation}
While the Hilbert space and the action of basic operators are much
simpler than in the full theory, their technical features are the
same: As already noted, the Hilbert space is
non-separable. Furthermore, only holonomies, not $c$ itself are
promoted to well-defined operators; it is impossible to derive an
operator for $c$ from the holonomy operators by differentiation since
the operator family $(e^{itc})_{t\in\R}$ is not continuous in $t$ at
$t=0$:
\begin{equation} \label{noncont}
\langle\mu|e^{itc}|\mu\rangle = \langle\mu|\mu+t\rangle =
\delta_{t,0}\,.
\end{equation}
Also as in the full theory, the flux operator $\hat{p}$ has a discrete
spectrum: its eigenstates $|\mu\rangle$ are normalizable (even though
a continuous range of eigenvalues is allowed).

The results (\ref{states}), (\ref{inner}), (\ref{c}) and (\ref{p}) can
also be used as axioms to define isotropic loop quantum cosmology,
without following the reduction from the full theory.  This may then
be seen as a quantization alternative to and, as we will see,
different from the Wheeler--DeWitt quantization (see also
[\onlinecite{BohrADM}] for a similar discussion in terms of ADM
variables).

In order to compare this basic framework with that of the
Wheeler--DeWitt quantization, we start with the formal Hilbert
spaces. In a Wheeler--DeWitt quantization one would simply employ
a Schr\"odinger quantization of the single degree of freedom $c$
such that the Hilbert space is $L^2(\R,dc)$ and the conjugate
momentum $p$ acts as a derivative operator. The loop quantization
is based on a different, inequivalent Hilbert space since it is
non-separable. It can be written as $L^2(\bar{\R}_{\rm
Bohr},d\lambda)$ where $\bar{\R}_{\rm Bohr}$ is the Bohr
compactification of the real line $\R$. It is a {\em compact}
Abelian group which contains $\R\subset\bar{\R}_{\rm Bohr}$ as a
dense subset. The measure $d\lambda$ of the isotropic loop Hilbert
space is the Haar measure on this compact group.

Thus, the Wheeler--DeWitt quantization and the loop quantization
are based on inequivalent Hilbert spaces such that they cannot be
isomorphic to each other. This is possible even with finitely many
degrees of freedom since the Weyl algebra $(e^{itc},
e^{isp})_{t,s\in\R}$ is not represented continuously in the loop
representation; see (\ref{noncont}). This assumption of the
Stone-von Neumann theorem, which is usually used to prove the
uniqueness of the quantum mechanical representation, is thus
violated. As a consequence we have the basic differences for the
operators: in the Wheeler--DeWitt quantization there is an
operator $\hat{c}$, which does not exist in the loop case, and the
operator $\hat{p}$ of the Wheeler--DeWitt quantization has a
continuous spectrum compared to the discrete one in the loop
quantization.

On the other hand, the loop quantization of the isotropic model
shares basic features with full loop quantum gravity. In the full
theory, the quantum configuration space is a compactification
$\bar{\mathcal A}$ of the classical space of connections
${\mathcal A}$ which is contained in $\bar{\mathcal A}$ as a dense
subset. This can be seen by analyzing the projective limit
construction sketched above.\cite{ALMMT} The situation in the
isotropic case is completely analogous: the isotropic space of
connections ${\mathcal A}_{\rm iso}\cong\R$ is extended and
compactified to the quantum configuration space $\bar{\mathcal
A}_{\rm iso}\cong\bar{\R}_{\rm Bohr}$ which contains ${\mathcal
A}_{\rm iso}$ as a dense subset. Here, the same mechanism occurs
just in a simpler context. As consequences we have both in the
full theory and in the isotropic case a non-separable Hilbert
space, no well-defined operators for connections, and flux
operators with discrete spectra. There is a difference, though,
namely that the isotropic flux operator has a continuous set of
eigenvalues even though its spectrum is discrete. This property
comes from a degeneracy manifest in the reduction procedure: the
label $\mu$ of an isotropic state $e^{i\mu c}$ does not
distinguish between the parameter length of an edge and the spin
when we compare it to a holonomy. There are technical subtleties
resulting from this issue that can be dealt with appropriately
(see [\onlinecite{Bohr}] for a detailed discussion of all these
concepts). Isotropic models, then, provide reliable tests of basic
and important features of the full theory. Calculations in these
models are certainly much simpler such that explicit results can
be obtained. This is already shown by the volume operator which is
rather complicated in the full theory while its spectrum can be
obtained explicitly in the isotropic context.\cite{cosmoII} Since
it plays an important role in constructing other operators such as
the Hamiltonian constraint,\cite{QSDI,QSDV} also the dynamical
equations become more treatable. We now turn to the physical
applications which can be derived in this way.

\section{Cosmology}

The dynamical equation of isotropic cosmology is the Friedmann
equation which is the same as the Hamiltonian constraint equation.  In
connection variables, the isotropic gravitational part of the
constraint reads
\begin{equation}
 H=-12\gamma^{-2}\kappa^{-1} (c(c-k)+(1+\gamma^2)k^2/4)
 \sqrt{|p|}
\end{equation}
with the gravitational constant $\kappa=8\pi G$ and the curvature
parameter $k$ which is zero for the spatially flat model and one for
the closed model. In order to relate this to the usual form of the
Friedmann equation, we need to know the relation between the basic
variables $(c,p)$ and the scale factor $a$. For the triad component,
one has $|p|=a^2$ ($p$ can be negative, indicating the orientation of
the triad), while the connection component is
$c=(k-\gamma\dot{a})/2$. If we insert this into the constraint
equation $H+H_{\rm matter}=0$ with some matter Hamiltonian, we obtain
in fact the Friedmann equation
\[
 3(\dot{a}^2+k^2)a=8\pi G a^3\rho(a)\,.
\]

\subsection{Difference Equation}

The quantization of the Hamiltonian constraint is rather complicated (see
[\onlinecite{Closed}] for the expression) since it is a composite
operator in terms of the basic ones, holonomies and fluxes. To
solve the constraint, we need to act on a general state
$|\psi\rangle = \sum_{\mu}\psi_{\mu}(\phi) |\mu\rangle$ where the
coefficients depend on matter fields which we collectively denote
as $\phi$. It is more convenient to write down the constraint
equation for the coefficients $\psi_{\mu}(\phi)$ directly, i.e.\
to work in a triad rather than connection representation, which we
can later compare to the Wheeler--DeWitt constraint. In this way,
our dynamical law becomes the difference equation
\begin{eqnarray} \label{disc}
 &&(V_{\mu+5}-V_{\mu+3}) e^{ik} \psi_{\mu+4}(\phi)- (2+\gamma^2k^2)
 (V_{\mu+1}-V_{\mu-1})\psi_{\mu}(\phi)\nonumber\\
 &&\hspace{1cm}+
 (V_{\mu-3}-V_{\mu-5}) e^{-ik} \psi_{\mu-4}(\phi)
 = -\frac{1}{6}\gamma^3\kappa\lP^2\hat{H}_{\rm
 matter}(\mu)\psi_{\mu}(\phi)
\end{eqnarray}
using the volume eigenvalues
\begin{equation}
 V_{\mu}=(\gamma\lP^2|\mu|/6)^{3/2}\,.
\end{equation}

A Wheeler--DeWitt quantization, in contrast, would result in a
differential equation which turns out to be a good approximation to
(\ref{disc}) at large volume, $|\mu|\gg1$, where the differences can
be Taylor expanded.\cite{SemiClass} At small $\mu$, however, the
discrete formulation is very different from the continuous one.  This
is crucial when we consider the fate of the classical singularity
which is not removed in the Wheeler--DeWitt quantization: generically,
curvatures and energy densities still diverge and the classical
singularity presents a boundary to the evolution (in some special
models one can argue for a more regular behavior in a different sense,
see e.g.\ [\onlinecite{Coule}]).  In the loop quantization, on the other
hand, it turns out that there is no singularity;\cite{Sing,IsoCosmo}
instead, the dynamical law provides a unique evolution from the wave
function at positive $\mu$ to its values at negative $\mu$.  This can
be seen directly when one starts with initial data at large positive
$\mu$ and uses (\ref{disc}) as a recurrence relation to evolve
backward toward the classical singularity at $\mu=0$. There are no
problems at first as long as the lowest order coefficient,
$V_{\mu-3}-V_{\mu-5}$, in the difference equation is non-zero. For
$\mu=4$, however, this coefficient is zero which means that
$\psi_0(\phi)$ remains undetermined in this way. If this value would
then appear in the next stages of the recurrence, the evolution would
break down and the classical singularity would present a boundary also
for the quantum evolution. As one can see, this does not occur: also
for $\mu=0$ and $\mu=-4$, where $\psi_0$ would appear in (\ref{disc})
at first sight, it drops out thanks to vanishing coefficients. Here we
use the fact that quantum geometry Hamiltonians imply the special
feature\cite{InvScale} $\hat{H}_{\rm matter}(0)=0$ which is
independent of quantization ambiguities\cite{Ambig} and comes from a
general procedure in the full theory.\cite{QSDV} Thus, $\psi_0$ also
drops out in the matter part of the constraint equation and we can
follow the evolution completely to negative $\mu$. (For a classical
picture of the corresponding bounce in a closed model, see
[\onlinecite{BounceClosed}].)

The classical singularity, therefore, does not appear as a natural
starting point where one could pose initial conditions.  Nevertheless,
the issue of boundary conditions appears in a new light
now.\cite{DynIn,Essay} When following the evolution we noticed that
$\psi_0$ remains undetermined and we ignored the part of the
constraint where it dropped out the first time ($\mu=4$).  However,
this part of the constraint still has to be satisfied providing us
with a linear relation between $\psi_4$ and $\psi_8$.  This, in turn,
translates to a linear relation between the initial data we chose. In
the isotropic case this is just the right amount to specify the
gravitational wave function completely up to its norm (ignoring the
matter field dependence) as usually done in the Wheeler-DeWitt
quantization. In order to compare with Wheeler--DeWitt wave functions
we have to assume that the solution we obtain is almost constant at
large volume at least for positive $\mu$, i.e.\ that it does not have
wild oscillations at the Planck scale. This condition
(pre-classicality\cite{DynIn,FundamentalDisc}) ensures that the
solution has a continuum limit to be compared with a Wheeler--DeWitt
wave function. There are two independent such solutions which, when
using the linear relation, are reduced to a unique (up to norm) linear
combination. In this way, the dynamical law implies the boundary
condition ({\em dynamical initial condition}) which is not chosen
separately.

\subsection{Boundary Proposals}

Boundary proposals for the Wheeler--DeWitt quantization have been
studied for some time, mainly focusing on the
tunneling\cite{tunneling} and no-boundary proposals.\cite{nobound}
We now have a very different formulation in loop quantum cosmology
with a discrete evolution equation and derived boundary
conditions. Nevertheless, after evolving the wave function to
large volume, where the Wheeler--DeWitt quantization is a very
good approximation, one can compare different boundary proposals
and possible physical implications. For this purpose, we will only
consider positive $\mu$ and $p$ for the rest of this section.

\begin{figure}[ht]
\begin{center}
\includegraphics[width=12cm,height=10cm,keepaspectratio]{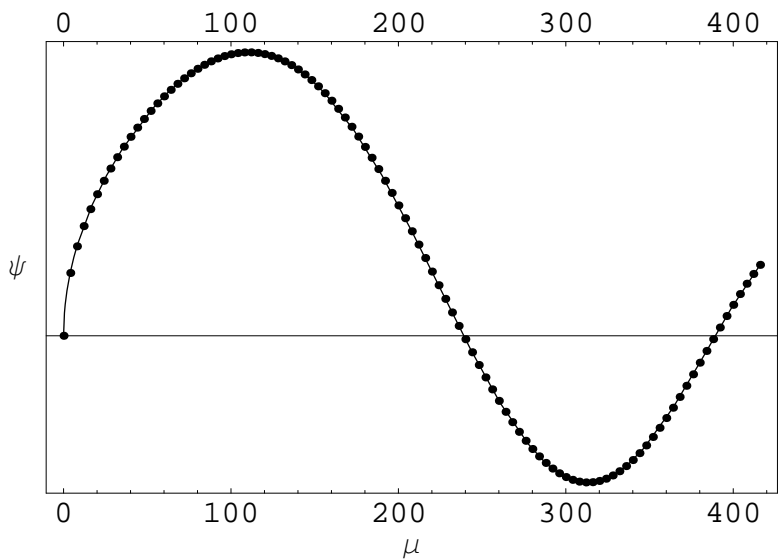}
\end{center}
\caption{Solution of the discrete equation for de Sitter space
with flat slices (dots), compared to a Wheeler--DeWitt solution
solving DeWitt's condition (solid line). For any other initial
condition, the continuous solution would diverge at $\mu=0$ and thus
deviate strongly from the discrete one.}
\label{deSitter}
\end{figure}

The first proposal is due to DeWitt who suggested that the wave
function should vanish at the classical singularity,\cite{DeWitt}
$\psi(0)=0$. The dynamical initial conditions in fact coincide
with this one for simple systems such as de Sitter space with flat
slices (see Fig.~\ref{deSitter}).\cite{DynIn} In more 
complicated cases, however, the discreteness at small volume implies
differences in the behavior of the wave functions. In this way,
the dynamical initial conditions even work for models where
DeWitt's condition would not be well-posed.\cite{Scalar}

Other boundary proposals have almost exclusively been studied for the
closed model, which is in particular true for the tunneling and
no-boundary proposals. In the formulation of the proposals, one
simply considers the closed model with a cosmological constant, where
there is a classically forbidden region between values of the scale
factor between zero and some minimum classically allowed value. 
There are detailed evaluations of their different implications, but the
central difference we are concerned with here is as follows. The
tunneling wave function starts with a large value at $a=0$ and
then decays exponentially in the classically forbidden region,
which is motivated as a tunneling process of the universe. After
that, the wave function becomes oscillating in the classically
allowed region. The no-boundary proposal, on the other hand, does
have a non-vanishing exponentially increasing contribution to the
wave function which will soon dominate. Thus, the wave function
starts with a small value at $a=0$ and grows much higher before
reaching the oscillatory phase. This different behavior is claimed
to have consequences for chaotic inflation when one computes the
probability distribution for the initial inflaton. This issue is
still disputed, but it is often said that the no-boundary proposal
prefers small initial values for the inflaton which would result
in an insufficient amount of inflation.

\begin{figure}[ht]
\begin{center}
\includegraphics[width=12cm,height=10cm,keepaspectratio]{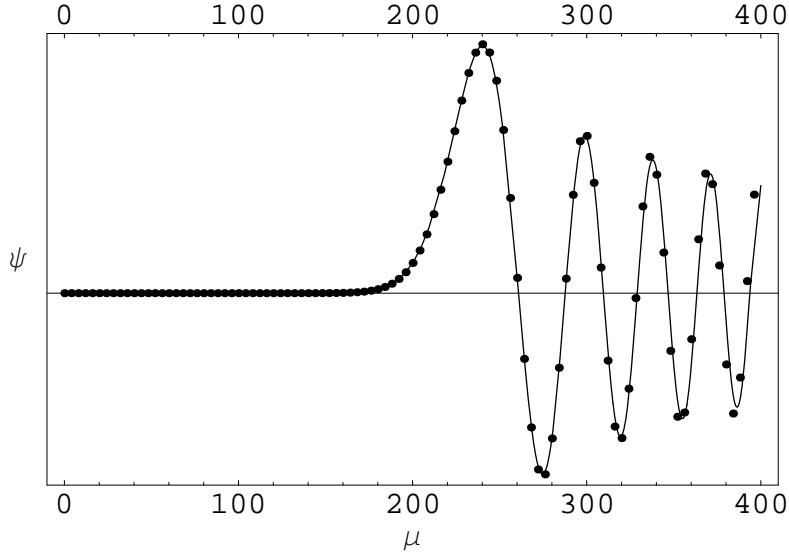}
\end{center}
\caption{Solution of the discrete equation (\ref{disc}) for
  de Sitter space with positive curvature spatial slices (dots) compared
  to a solution of the continuous Wheeler--DeWitt equation 
  which is regular at $\mu=0$ (solid line).}
\label{WaveFunc}
\end{figure}

\begin{figure}[ht]
\begin{center}
\includegraphics[width=12cm,height=10cm,keepaspectratio]{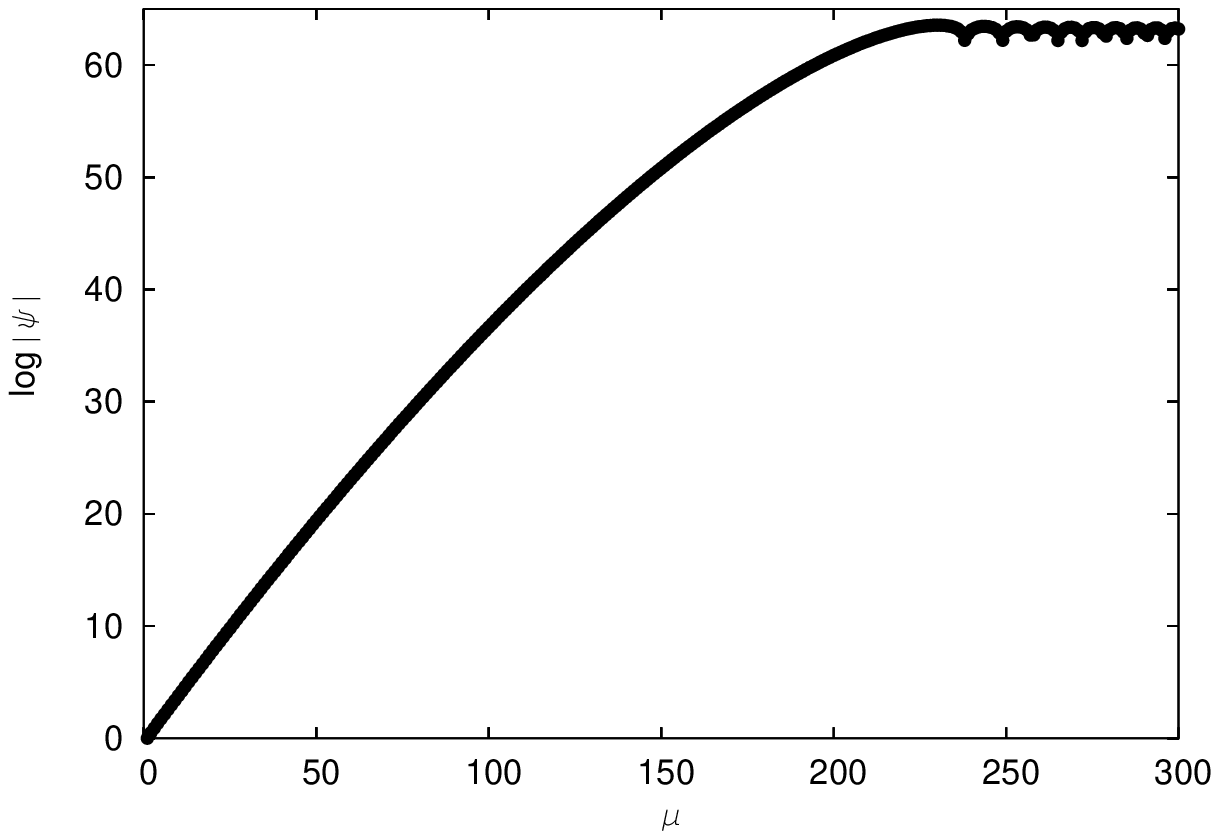}
\end{center}
\caption{Logarithm of the absolute value of the wave function. In the
  classically forbidden region the wave function grows by more than 60
  orders of magnitude.}
\label{WaveGamLog}
\end{figure}

In order to see how the loop wave function fits into this
situation it is sufficient to note that the tunneling wave
function is very special. It requires that the exponentially
growing part does not contribute to the wave function which can be
achieved only in this one very special case. Any other proposal
would have a non-vanishing contribution of the growing mode and
would soon differ from the tunneling wave function. Thus, any
proposal which is not tailored to be exactly the tunneling one,
would be much closer to the no-boundary condition at large volume.
This is in particular true for the loop wave function with its
dynamical initial condition. A numerical analysis
(Figs.~\ref{WaveFunc} and \ref{WaveGamLog}) shows that this wave
function also increases exponentially in the classically forbidden
region and thus is closer to the no-boundary proposal. Whether or
not there are physically significant differences to the
no-boundary wave function also at large volume remains to be
studied.

\begin{figure}[ht]
\begin{center}
\includegraphics[width=12cm,height=10cm,keepaspectratio]{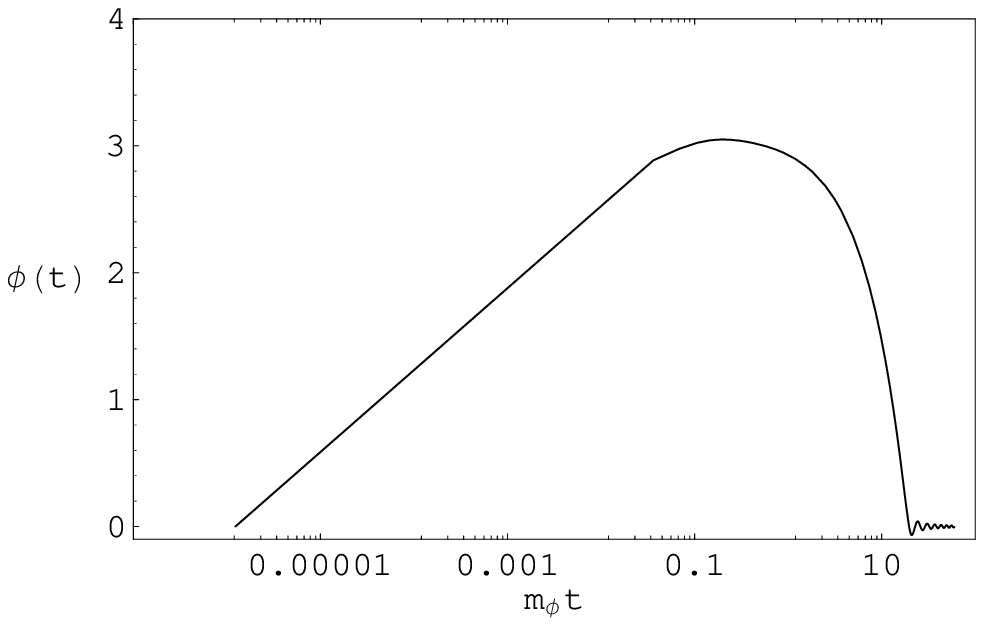}
\end{center}
\caption{Classical evolution of an inflaton field in a quadratic
potential with quantum modifications to the kinetic term of the matter
Hamiltonian. The quantum modifications help drive the scalar field
at early stages after which slow roll inflation can occur.}
\label{ScalarField}
\end{figure}

This conclusion may seem bad if the no-boundary proposal does in fact
not provide enough inflation. However, when we consider the evolution
of a scalar or other matter field coupled to the discrete quantum
geometry, there exist small-volume quantum modifications to the
kinetic term of the matter Hamiltonian which automatically lead to
inflation.\cite{Inflation} The modifications arise from the
quantization of the inverse volume which removes the classical
divergence In the situation of chaotic inflation where one couples an
inflaton $\phi$, the modified inflaton dynamics implies that $\phi$ is
driven up its potential at early stages (see Fig. \ref{ScalarField}).
Before scales much larger than the Planck length are reached, the
inflaton turns around and enters a slow-roll phase of chaotic
inflation. Thus, the traditional considerations of probability
distributions for the inflaton are not even needed here since there is
a dynamical mechanism which results in large inflaton values. This
occurs for a wide range of initial conditions with the initial value
of the scalar field and its conjugate momentum being constrained by
the uncertainty principle.\cite{InflationWMAP} The particular behavior
may even have observable consequences since the slow-roll condition is
violated around the turning point, i.e.\ at early stages of the
slow-roll phase translating into large current
scales.\cite{InflationWMAP}

\section{Conclusions}

The overall picture we have seen from loop quantum cosmology is
very different from the traditional Wheeler--DeWitt quantization,
and its structure is much richer. There are many phenomenological
effects: A cut-off of classically diverging
densities,\cite{InvScale,Ambig} which is well-studied in flat
models, implies inflation\cite{Inflation} with possibly observable
effects\cite{InflationWMAP} when coupled to chaotic inflation.
Also without coupling an inflaton, the universe would still expand
with positive acceleration. Whether or not this is physically
viable is undecided at this time since the usual perturbation
theory becomes inapplicable.\cite{InflationWMAP}

In the closed model with its classically forbidden region there are
even more diverse phenomenological possibilities which have not yet
been studied in detail. In this case, not just densities, or extrinsic
curvature, experience a cut-off by quantum geometry effects, but also
the intrinsic curvature can become suppressed at small
volumes.\cite{Closed} There the closed model behaves similarly to the
flat model with intrinsic curvature close to zero and a classically
allowed region can be created near the singularity. It remains to be
seen what physical consequences these effects have as the classically
allowed regions lies deep within the quantum geometry regime at small
volume.

\begin{figure}[ht]
\begin{center}
\includegraphics[width=12cm,height=10cm,keepaspectratio]{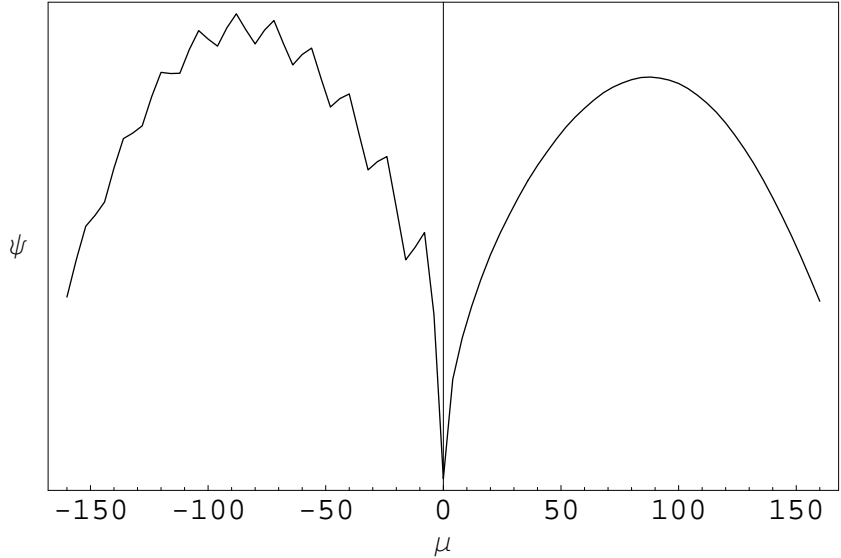}
\end{center}
\caption{Solution to the more complicated discrete equation modeled more
closely to the full theory for de Sitter space with flat spatial slices. 
The Planck scale oscillations persist
for large negative values of $\mu$. The simpler evolution equation
(\ref{disc}) would give a wave function symmetric about $\mu=0$.}
\label{FlatLorentz}
\end{figure}

Other open issues are concerned with details of the evolution through
the classical singularity. Since this involves the dynamics deep
inside the Planck regime, results are very sensitive to quantization
choices. This may be used to find restrictions among possible choices,
but there are also many different cases which have to be studied. The
equation (\ref{disc}) used here implies a symmetric solution which
leads to equal behavior at positive or negative $\mu$. But there is a
more complicated version\cite{IsoCosmo} which is closer to the
constraint in the full theory.\cite{QSDI} The properties discussed
here are insensitive to which one of those two forms for the
constraint is used. But if we compare the wave function at positive
and negative $\mu$ obtained from the more complicated constraint, then
there will usually be significant small-scale oscillations at one side
if they are constrained to be small at the other (see
Fig.~\ref{FlatLorentz}).\cite{DynIn} These oscillations arise because
the more complicated constraint is of higher order in internal time
than the simpler one (also $\psi_{\mu\pm8}$ are involved).  Being
discrete, the evolution equation may contain coefficients which do not
behave smoothly especially near the classical singularity. Thus
passing through the singularity a smooth wave function can have the
Planck scale frequency modes excited whose amplitudes can be very
large in some models.\cite{ScalarLorentz} Particular quantizations
choices can be favored if they provide an evolution equation with
smooth coefficients, which may be exploited to reduce the freedom in
the full quantization.

From the continuum picture point of view, the freedom for solutions to
the discrete equation is reduced compared to the Wheeler--DeWitt
equation. Since one has to require the existence of a continuum limit
(or at least a good continuum approximation\cite{FundamentalDisc}) for
the comparison, equation (\ref{disc}) is reduced to second order (even
for the more complicated version mentioned in the preceding
paragraph). The dynamical initial condition then leads to a reduction
in the number of free parameters as seen before. From the discrete
point of view, however, there are many more solutions (infinitely many
since $\mu$ is continuous) with Planck scale oscillations. Their role
and physical interpretation are open issues, and it is expected that
an understanding requires the so far unknown physical inner product
and the consideration of quantum observables. Tentative ideas in this
direction can be found in [\onlinecite{cosmoIV,Golam}], and further
investigations are in progress.

There are promising possibilities for future developments.  It is
possible to extend the results to less symmetric models, which has
already been completed for anisotropic but still homogeneous
ones\cite{HomCosmo,Spin} resulting in new physical
effects.\cite{NonChaos} For cosmological considerations it is also
important to add inhomogeneities at least as perturbations or in
midisuperspace models. This will be done in the future in order to
complete the picture provided by loop quantum cosmology, but even
before these technical developments much remains to be studied, as
indicated above, even in isotropic models.

\smallskip

M.~B.\ is grateful to C.~Kiefer for an invitation to a talk
at the Xth Marcel Grossmann meeting, July 20--26, 2003, Rio de
Janeiro, on which this paper is based.  This work was supported in
part by NSF grant PHY00-90091 and the Eberly research funds of Penn
State.

\end{document}